\documentclass[a4paper,onecolumn,11pt,accepted=2025-03-03]{quantumarticle}
\pdfoutput=1
\usepackage[utf8]{inputenc}
\usepackage[english]{babel}
\usepackage[T1]{fontenc}
\usepackage{hyperref}
\usepackage{braket}
\usepackage{graphicx}
\usepackage{enumitem}
\usepackage{booktabs}
\usepackage[frozencache]{minted}
\usepackage{qcircuit}
\usepackage{subcaption}
\usepackage{multirow}
\usepackage{mathtools}
\usepackage{nicematrix}
\usepackage[numbers,sort&compress]{natbib}

\begin{document}

\title{Optimizing Gate Decomposition for High-Level Quantum Programming}

\author{Evandro C. R. Rosa}
\orcid{0000-0002-8197-9454}
\email{evandro.crr@posgrad.ufsc.br}

\author{Eduardo I. Duzzioni}
\orcid{0000-0002-8971-2033}

\author{Rafael de Santiago}
\orcid{0000-0001-7033-125X}

\affiliation{Universidade Federal de Santa Catarina, Grupo de Computação Quântica, 88040-900 Florianópolis, SC, Brazil}

\begin{abstract}
    This paper presents novel methods for optimizing multi-controlled quantum gates, which naturally arise in high-level quantum programming. Our primary approach involves rewriting $U(2)$ gates as $SU(2)$ gates, utilizing one auxiliary qubit for phase correction. This reduces the number of CNOT gates required to decompose any multi-controlled quantum gate from $O(n^2)$ to at most $32n$. Additionally, we can reduce the number of CNOTs for multi-controlled Pauli gates from $16n$ to $12n$ and propose an optimization to reduce the number of controlled gates in high-level quantum programming. We have implemented these optimizations in the Ket quantum programming platform and demonstrated significant reductions in the number of gates. For instance, for a Grover's algorithm layer with 114 qubits, we achieved a reduction in the number of CNOTs from 101,252 to 2,684. This reduction in the number of gates significantly impacts the execution time of quantum algorithms, thereby enhancing the feasibility of executing them on NISQ computers.
\end{abstract}

\section{Introduction}

Quantum computing is an emerging computational paradigm that utilizes the principles of quantum mechanics to process information. Quantum computers can store and manipulate data using superposition and entanglement, enabling them to potentially solve certain problems more efficiently than any classical supercomputer. This novel approach to computation has impacts across various industries, including finance~\cite{hermanQuantumComputingFinance2023}, logistics~\cite{osabaSystematicLiteratureReview2022}, chemistry~\cite{caoQuantumChemistryAge2019}, cryptography~\cite{gidneyHowFactor20482021}, and artificial intelligence~\cite{biamonteQuantumMachineLearning2017}.

Despite achieving the milestone of quantum advantage~\cite{preskillQuantumComputingEntanglement2012}, where problems that would take days to solve on classical supercomputers can be solved in seconds~\cite{aruteQuantumSupremacyUsing2019,madsenQuantumComputationalAdvantage2022,zhongQuantumComputationalAdvantage2020,wuStrongQuantumComputational2021,zhongPhaseProgrammableGaussianBoson2021}, quantum computing technology is not yet ready to address industrial-scale problems in production environments. We are currently in the NISQ era~\cite{preskillQuantumComputingNISQ2018}, characterized by Noisy Intermediate-Scale Quantum computers containing dozens to a few hundred quantum bits, which are limited by the error rates of current systems.

Several companies and startups are currently developing quantum computers using various technologies, including superconductors~\cite{kjaergaardSuperconductingQubitsCurrent2020}, trapped ions~\cite{eganFaulttolerantControlErrorcorrected2021}, neutral atoms~\cite{henrietQuantumComputingNeutral2020}, quantum dots~\cite{yonedaQuantumdotSpinQubit2018}, and diamond NV centers~\cite{childressDiamondNVCenters2013}. While each technology possesses its unique capabilities, they all share the same programming paradigm: utilizing quantum bits, or qubits, to store information and quantum gates to perform computations. Although quantum computers offer a universal set of quantum gates capable of executing any computation, there are two constraints for high-level quantum programming. Firstly, these gates are limited to operating on one or two qubits. Secondly, there exists only a finite set of quantum gates available.

Regarding the latter constraint, we can always map a one- or two-qubit gate into a combination of others, meaning that despite the limited options of quantum gates, there exists a function, which can be calculated beforehand, to perform this transformation. However, for the former case, decomposing an arbitrary $n$-qubit gate into a sequence of one- and two-qubit gates takes exponential time relative to the number of qubits~\cite{itenQuantumCircuitsIsometries2016,itenIntroductionUniversalQCompiler2019}. Fortunately, there exists a set of multi-qubit quantum gates that are useful for high-level quantum programming, along with known efficient decomposition algorithms.

A finite set of single-qubit quantum gates, coupled with the ability to add control qubits to these gates' applications, is sufficient to enable high-level quantum programming instructions~\cite{darosaKetQuantumProgramming2022}, including support for quantum subroutines. These single-qubit gates must enable the preparation of a qubit into any arbitrary quantum state, global phase considered, as we will explore later in this article.

A comprehensive set of single-qubit quantum gates that encompass most quantum algorithms includes the Pauli gates, the Hadamard gate, the phase shift gate, and the parametrized rotation gates. For any single-qubit gate with $n$ qubits of control, there exists an efficient decomposition algorithm with a time complexity of $O(n^2)$~\cite{dasilvaLineardepthQuantumCircuits2022}, but specifically for multi-controlled rotation gates, there exists an $O(n)$ decomposition algorithm~\cite{valeCircuitDecompositionMulticontrolled2024}. The rotation gates belong to the set of $SU(2)$ gates, which is a subset of $U(2)$ containing all single-qubit gates.

In this paper, we propose a series of optimizations for multi-control single-target quantum gate decomposition, implementing and evaluating the results in the Ket quantum programming platform~\cite{darosaKetQuantumProgramming2022}. The contributions of the paper are:
\begin{enumerate}[label=\textbf{C.\arabic*}]
    \item \label{C1} A schema to describe any multi-controlled $U(2)$ gate as multi-controlled $SU(2)$ gates, leveraging the $O(n)$ decomposition with the trade-off of adding a single auxiliary qubit. This allows programmers to describe their applications in terms of $U(2)$ without incurring the $O(n^2)$ decomposition cost.
    \item \label{C2} Reducing the number of CNOTs for $n$-controlled $SU(2)$ gates from at most $20n$ to $16n$, a result previously limited to $SU(2)$ gates with at least one real diagonal~\cite{valeCircuitDecompositionMulticontrolled2024}.
    \item \label{C3} Since that the auxiliary qubit in our proposed optimization for multi-controlled $U(2)$ gates is in the state $\ket{0}$, we can reduce the number of CNOTs to decompose Pauli gates from at most $16n$~\cite{itenQuantumCircuitsIsometries2016} to $12n$.
    \item \label{C4} Presenting methods to leverage reversible quantum computation to reduce the number of controlled gates in Ket's \texttt{with around} high-level quantum programming statement.
    \item \label{C5} Implementing all the optimizations in the Ket quantum programming platform\footnote{\url{https://quantumket.org}}, allowing quantum programs to take advantage of these techniques without any changes to their quantum code.
\end{enumerate}

The decomposition of multi-qubit gates represents a crucial first step in preparing a quantum application for execution on a quantum computer~\cite{sivarajahT|ketRetargetableCompiler2021,nguyenExtendingHeterogeneousQuantumClassical2020,mccaskeyXACCSystemlevelSoftware2020}. Consequently, optimizing this process significantly impacts the overall quantum compilation workflow. After decomposing quantum gates into one- and two-qubit gates, the next stage involves adjusting the two-qubit gates to align with the connectivity constraints of the target quantum computer~\cite{liTacklingQubitMapping2019,huaQASMTransQASMQuantum2023,willeMQTQMAPEfficient2023}, a procedure referred to as quantum circuit mapping. Thus, the optimization introduced in this article, which reduces the number of quantum gates, directly influences the efficiency of quantum circuit mapping. This mapping problem is known to be NP-Hard~\cite{siraichiQubitAllocation2018,boteaComplexityQuantumCircuit2021}.

For the remainder of this paper, we will structure our discussion as follows: In Section~\ref{sec:high-level}, we will discuss how multi-controlled $U(2)$ gates enable the construction of quantum subroutines in high-level quantum programming, while also exploring how high-level quantum programming can be used to reduce the number of controlled operations. In Section~\ref{sec:decomposition}, we will present the decomposition algorithms for multi-controlled $SU(2)$ and Pauli gates. Following that, in Section~\ref{sec:replace}, we will demonstrate how to rewrite any $U(2)$ gate in terms of $SU(2)$ gates, allowing us to leverage the $O(n)$ decomposition for any multi-controlled single-target quantum gate. Subsequently, in Section~\ref{sec:results}, we will present numerical results showcasing the impact of this rewrite and other optimizations on the classical and quantum runtime. Finally, we conclude in Section~\ref{sec:conclusion}, summarizing our findings and discussing future work.

\section{High-Level Quantum Programming}
\label{sec:high-level}

At the lowest level, a quantum program consists of a sequence of radio frequency pulses~\cite{stefanazziQICKQuantumInstrumentation2022}, analogous to a binary program for a classical computer. Consequently, programming at this level is complex and heavily dependent on the target hardware. Therefore, quantum computers typically provide a set of one and two-qubit gates, where these gates have been pre-calibrated and have a known pulse format~\cite{kimmelRobustCalibrationUniversal2015,huangCalibratingSinglequbitGates2023}. This set of gates forms the equivalent of an assembly language for a quantum computer, where programs are still hardware-dependent, but abstraction allows for the development and testing of simple quantum programs.

Similar to a classical Instruction Set Architecture (ISA)~\cite{pattersonComputerOrganizationDesign2018} with a limited number of registers, quantum computers have a finite number of qubits, and the connectivity of these qubits limits which qubits can operate together in two-qubit gates. Therefore, the level of abstraction provided by a quantum assembly language is not suitable for the development of complex quantum applications. This necessitates the use of high-level quantum programming platforms such as Qiskit~\cite{javadi-abhariQuantumComputingQiskit2024}, Cirq~\cite{cirqdevelopersCirq2024}, ProjectQ~\cite{steigerProjectQOpenSource2018}, PennyLane~\cite{bergholmPennyLaneAutomaticDifferentiation2018}, Q\#\cite{svoreEnablingScalableQuantum2018}, and Ket~\cite{darosaKetQuantumProgramming2022}. At this level, we transition away from pulse definitions and hardware constraints towards the mathematical formalism of quantum mechanics. Using a quantum compiler, we can take a high-level quantum application and execute it on a quantum computer.

Table~\ref{tab:quantum_gates} presents a gate set commonly used in many quantum algorithms. Additionally, controlled versions of these gates are also utilized. For example, the CNOT gate, short for Controlled-NOT, is a Pauli~$X$ gate (also known as the NOT gate) with a control qubit. The matrix representation of the CNOT gate is as follows:
\begin{equation}
    \text{CNOT} = \begin{bNiceMatrix}
        1 & 0 & 0 & 0 \\
        0 & 1 & 0 & 0 \\
        0 & 0 & 0 & 1 \\
        0 & 0 & 1 & 0 \\
    \end{bNiceMatrix}.
\end{equation}

\begin{table}[htbp]
    \centering
    \caption{Commonly used single-qubit quantum gates. These gates form the foundational set implemented in Ket's runtime library, Libket. All other gates in Ket are composed using controlled versions of these base gates.}
    \label{tab:quantum_gates}
    \footnotesize
    \begin{tabular}[t]{cc}
        \toprule
        \textbf{Gate Name} & \textbf{Matrix Representation}                                                              \\
        \midrule
        Pauli~$X$          & $X= \begin{bNiceMatrix} 0 & 1 \\ 1 & 0 \end{bNiceMatrix}$ \vspace{5pt}                      \\
        Pauli~$Y$          & $Y= \begin{bNiceMatrix} 0 & -i \\ i & 0 \end{bNiceMatrix}$ \vspace{5pt}                     \\
        Pauli~$Z$          & $Z= \begin{bNiceMatrix} 1 & 0 \\ 0 & -1 \end{bNiceMatrix}$ \vspace{5pt}                     \\
        Hadamard           & $H = \frac{1}{\sqrt{2}} \begin{bNiceMatrix} 1 & 1 \\ 1 & -1 \end{bNiceMatrix}$ \vspace{5pt} \\
        \bottomrule
    \end{tabular}
    \begin{tabular}[t]{cc}
        \toprule
        \textbf{Gate Name} & \textbf{Matrix Representation}                                                                                                            \\
        \midrule
        Phase              & $P(\theta) = \begin{bNiceMatrix} 1 & 0 \\ 0 & e^{i\theta} \end{bNiceMatrix}$ \vspace{5pt}                                                 \\
        $X$ Rotation       & $R_X(\theta) = \begin{bNiceMatrix} \cos(\theta/2) & -i\sin(\theta/2) \\ -i\sin(\theta/2) & \cos(\theta/2) \end{bNiceMatrix}$ \vspace{5pt} \\
        $Y$ Rotation       & $R_Y(\theta) = \begin{bNiceMatrix} \cos(\theta/2) & -\sin(\theta/2) \\ \sin(\theta/2) & \cos(\theta/2) \end{bNiceMatrix}$ \vspace{5pt}    \\
        $Z$ Rotation       & $R_Z(\theta) = \begin{bNiceMatrix} e^{-i\theta/2} & 0 \\ 0 & e^{i\theta/2} \end{bNiceMatrix}$ \vspace{5pt}                                \\
        \bottomrule
    \end{tabular}
\end{table}

A Controlled-$U$ gate, where $U$ is a unitary $2\times2$ matrix, is represented by the following matrix block form:
\begin{equation}
    \text{Controlled-}U = \begin{bNiceMatrix}
        I          & \mathbf{0} \\
        \mathbf{0} & U
    \end{bNiceMatrix},
\end{equation}
where $I$ represents an $2\times2$ identity matrix, and $\mathbf{0}$ signifies a $2\times2$ null matrix.

Controlled-gates operate with both a target qubit and a control qubit. The operation is applied to the target qubit only if the control qubit is in the state $\ket{1}$\footnote{While the control state of a qubit is typically $\ket{1}$, it can be any basis state. For simplicity, this paper assumes all control qubits are in the state $\ket{1}$.}. This notation can be extended for $n$-control qubits as
\begin{equation}
    C^nU = \sum_{k=0}^{2^n-2} {\ket{k}}{\bra{k}}\otimes I + {\ket{2^n{-1}}}{\bra{2^n{-1}}}\otimes U = \overbrace{ \begin{bNiceMatrix}
            1      & 0      & \Cdots  & 0      & 0       & 0       \\
            0      & 1      & \Cdots  & 0      & 0       & 0       \\
            \Vdots & \Vdots & \Ddots & \Vdots & \Vdots  & \Vdots  \\
            0      & 0      & \Cdots & 1      & 0       & 0       \\
            0      & 0      & \Cdots & 0      & U_{0,0} & U_{0,1} \\
            0      & 0      & \Cdots & 0      & U_{1,0} & U_{1,1} \\
        \end{bNiceMatrix}}^{2^{n+1}\times2^{n+1}},
    \label{eq:cnu}
\end{equation}
where $k$ is an integer, with $\ket{k}$ representing an $n$-qubit state. The gate $U$ is applied to the target qubit only if all $n$ control qubits are in the $\ket{1}$ state.

In quantum programming, integrating control qubits into quantum operations facilitates the straightforward expression of numerous quantum computations. It enables the description of various quantum subroutines, such as Grover's diffusion, a pivotal step in Grover's quantum search algorithm~\cite{groverFastQuantumMechanical1996}, and the Quantum Fourier Transformation, utilized in Shor's quantum factorization algorithm~\cite{shorPolynomialTimeAlgorithmsPrime1997}.

Figure~\ref{fig:ket_grover_op} illustrates the implementation of Grover's diffusion operator in the Ket quantum programming platform. This example highlights the use of high-level constructions. The \texttt{with around} instruction takes advantage of the quantum circuit's reversibility, reducing the number of instructions needed to describe a quantum circuit that starts with a gate and ends with the inverse of that same gate. In Figure~\ref{fig:grover_op}, we present the circuit for a 3-qubit Grover's diffusion. Notably, each qubit line begins with a Hadamard followed by a Pauli~$X$ gate ($XH$) and ends with its inverse ($(XH)^\dagger = HX$). The use of the \texttt{with around} statement minimizes the number of high-level quantum programming instructions needed for this implementation. Also, in the middle of the Grover's diffusion circuit, we observe a multi-controlled Pauli~$Z$ gate. In Ket, this controlled-gate can be implemented by placing a Pauli~$Z$ gate within a \texttt{with control} scope, which effectively adds control qubits to any quantum gate called inside.

\begin{figure}[htbp]
    \centering

    \hfill
    \begin{subfigure}[b]{.55\textwidth}
        \centering
        \begin{minipage}{.8\linewidth}
            \begin{minted}[frame=lines,fontsize=\small,linenos,breaklines]{py}
def grover_diffusion(qubits):
    with around(cat(H, X), qubits):
        with control(qubits[:-1]):
            Z(qubits[-1])
    \end{minted}
        \end{minipage}
        \caption{Grover's diffusion operator implemented using Ket.}
        \label{fig:ket_grover_op}
    \end{subfigure}
    \hfill
    \begin{subfigure}[b]{0.40\textwidth}
        \centering
        \begin{equation*}
            \Qcircuit @C=1em @R=1em {
             & \gate{H} & \gate{X} & \ctrl{2} & \gate{X} & \gate{H} & \qw \\
             & \gate{H} & \gate{X} & \ctrl{1} & \gate{X} & \gate{H} & \qw \\
             & \gate{H} & \gate{X} & \gate{Z} & \gate{X} & \gate{H} & \qw \\
            }
        \end{equation*}
        \caption{Grover's diffusion operator for 3 qubits.}
        \label{fig:grover_op}
    \end{subfigure}
    \hfill

    \caption{Grover's diffusion operator.}
\end{figure}

Figure~\ref{fig:ket_prepare} illustrates a state preparation algorithm implemented using the Ket quantum programming platform. The \texttt{prepare} function serves as a quantum gate, preparing a set of qubits into a quantum state encoded in two lists: one representing the measurement probabilities and the other encoding the phase of the state's amplitude probability. In this code snippet, lines 17 and 19 recursively invoke the \texttt{prepare} function, stacking control qubits with each call as they are nested within the scope of the \texttt{with control} in lines 16 and 18.

It is worth noting that although only single quantum gate applications are explicitly called in lines 9, 11-13, and 15, the recursive controlled call results in multi-controlled $R_Y$ and multi-controlled $P(\theta)$ gates, as depicted in Figure~\ref{fig:circuit_prepare}. Additionally, it's noteworthy that due to optimizations implemented in the \texttt{with around} statement, Ket does not add controls for the Pauli~$X$ gates. This optimization, which will be further elucidated in Subsection~\ref{subsec:opwitharound}, is also a contribution of this work.

\begin{figure}[htbp]
    \centering
    \begin{minipage}{.5\linewidth}
        \begin{minted}[frame=lines,fontsize=\small,linenos,breaklines]{py}
def prepare(
    qubits: Quant,
    prob: ParamTree | list[float | int],
    amp: list[float] | None = None,
):
    if not isinstance(prob, ParamTree):
        prob = ParamTree(prob, amp)
    head, *tail = qubits
    RY(prob.value, head)
    if prob.is_leaf():
        with around(X, head):
            PHASE(prob.phase0, head)
        PHASE(prob.phase1, head)
        return
    with around(X, head):
        with control(head):
            prepare(tail, prob.left)
    with control(head):
        prepare(tail, prob.right)
        \end{minted}
    \end{minipage}

    \caption{Recursive function for quantum state preparation using Ket. This implementation of the \texttt{ParamTree} class (detailed in Figure~\ref{fig:ket_param_tree}) features a tree graph structure. The algorithm traverses this tree graph, using node information to adjust measurement probabilities with $R_Y$ gates, and leaf node values to adjust the complex phase of the probability amplitude with Phase gates.}
    \label{fig:ket_prepare}
\end{figure}

The construction illustrated in Figure~\ref{fig:ket_prepare}, which allows calling a controlled function, is present in many high-level quantum programming platforms~\cite{darosaKetQuantumProgramming2022,svoreEnablingScalableQuantum2018}. It serves as an example of how controlled quantum operations can appear indirectly. As mentioned before, quantum computers do not implement these operations directly. Instead, algorithms exist to decompose these operations into one and two quantum gates. However, some gate decompositions are more efficient than others. For instance, an arbitrary $C^nU$ gate, such as the multi-controlled $P(\theta)$, decomposition results in a linear-depth circuit but has a decomposition time complexity of $O(n^2)$~\cite{dasilvaLineardepthQuantumCircuits2022}. On the other hand, if $U \in SU(2)$, such as the multi-controlled $R_Y$, the time complexity of the decomposition is linear~\cite{valeCircuitDecompositionMulticontrolled2024}. In this paper, we present an optimization that allows using the more efficient $SU(2)$ decomposition for any $C^nU$ gates. In the next sections, we illustrate how to perform the decomposition, and in section~\ref{sec:replace}, we present this decomposition optimization.

\begin{figure}[htbp]
    \includegraphics[width=\linewidth]{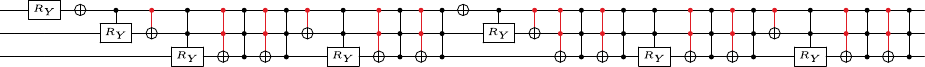}

    \caption{Quantum circuit resulting from executing the code from Figure~\ref{fig:ket_prepare} with three qubits. The red lines represent the control qubits that were optimized out in the \texttt{with around} statements, where multi-controlled Pauli~$X$ gates were replaced by Pauli~$X$ gates.}
    \label{fig:circuit_prepare}
\end{figure}

\subsection{Reducing the Number of Controlled Gates}
\label{subsec:opwitharound}

Another optimization proposed in this article is reducing the number of controlled operations in the \texttt{with around} construction (\ref{C4}). This statement, implemented in Ket, takes two unitary operations as arguments: the first is the \texttt{around} function arguments, and the second is the new code scope created in the next line, respectively, corresponding to the unitaries $A$ and $B$. For example, in lines 11 and 12 of the code in Figure~\ref{fig:ket_prepare}, the unitary $A$ is the application of the Pauli~$X$ gate, and the unitary $B$ is the application of the Phase gate.

The resulting unitary from a \texttt{with around} statement is $A^\dagger BA$. Prior to this optimization proposal, calling a \texttt{with around} statement with control qubits resulted in the unitary $(C^nA^\dagger)\cdot(C^nB)\cdot(C^nA)$, effectively adding control qubits to all operations. However, we can optimize this construction by adding control qubits only to the $B$ operation, resulting in $A^\dagger\cdot(C^nB)\cdot A$. The intuition behind this optimization is that if the control qubits are in the control state (all qubits in $\ket{1}$), then the entire unitary $A^\dagger BA$ will be applied to the target qubits. Conversely, if the control qubits are not in the control state, the $B$ unitary will not be applied, resulting in the application of $A^\dagger A = I$, which is equal to the identity operation, a no-operation.

With this optimization, we can replace all the controlled-NOT gates in the code from Figure~\ref{fig:ket_prepare}, which would otherwise result from the recursive controlled calls of lines 11 and 15, with simple NOT gates.  As shown in Figure~\ref{fig:circuit_prepare}, with only 3 qubits, this represents a reduction from 30 to 16 controlled gates. This is one example of how high-level quantum programming can result not only in fewer lines of code but also in more efficient code.

\section{Quantum Gate Decomposition}
\label{sec:decomposition}

Quantum gate decomposition is a crucial step in translating high-level quantum programming instructions into operations that a quantum computer can execute. In this paper, we focus on the decomposition of multi-control single-target quantum gates, as described in Equation~\eqref{eq:cnu}. Specifically, we examine the gates listed in Table~\ref{tab:quantum_gates}, although our results can be applied to any single-qubit unitary operation. To the best of our knowledge, the state-of-the-art in quantum gate decomposition is represented by the works of \citet{itenQuantumCircuitsIsometries2016}, \citet{valeCircuitDecompositionMulticontrolled2024}, and \citet{dasilvaLineardepthQuantumCircuits2022}. These works propose decompositions that result in linear-depth quantum circuits, with the first two limited to Pauli gates and $SU(2)$ gates, respectively, and the last one applicable to any $U(2)$ gate. For gates that are $2\times2$ special unitary operations ($SU(2)$), the decomposition can be performed with only a linear number of CNOTs, while for a general $U(2)$ acting with controlled qubit, the number of CNOTs is $O(n^2)$ \cite{dasilvaLineardepthQuantumCircuits2022}. From Table~\ref{tab:quantum_gates}, it can be observed that the rotation gates belong to $SU(2)$. For other quantum gates, with the exception of Pauli gates, the decomposition results in a quadratic number of CNOTs gates.

\begin{figure}
    \centering
    \includegraphics{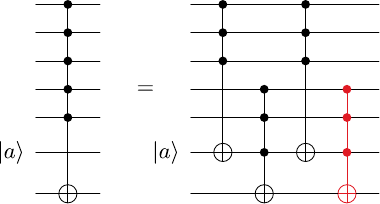}
    \caption{Multi-controlled Pauli~$X$ gate decomposition using a dirty auxiliary qubit. If the auxiliary qubit $\ket{a}$ is in the state $\ket{0}$, the last controlled operation in red can be discarded.}
    \label{fig:mxc_aux}
\end{figure}

For the Pauli~$X$ gate, \citet{barencoElementaryGatesQuantum1995} proposed a circuit to decompose a $k$-controlled Pauli~$X$ gate using $k-2$ dirty\footnote{These qubits can be in any state.} auxiliary qubits. This decomposition utilizes a chain of Toffoli gates ($C^2X$). Improving on this, \citet{itenQuantumCircuitsIsometries2016} proposed using an approximate Toffoli gate that requires only 3 CNOTs instead of 6, as in the exact case. This results in a decomposition that requires at most $4n$ CNOTs, where $n$ is the number of control qubits. Leveraging this optimization, \citet{itenQuantumCircuitsIsometries2016} further proposed a decomposition for the $n$-controlled Pauli~$X$ gate that requires at most $16n$ CNOTs and a single dirty auxiliary qubit. This method breaks the $n$-controlled Pauli~$X$ into four $\tfrac{n}{2}$-controlled Pauli~$X$ decompositions, as illustrated in Figure~\ref{fig:mxc_aux}, allowing the qubits that are not used in the multi-controlled Pauli~$X$ gate to be used as auxiliaries.

For the multi-controlled Pauli~$X$ gate decomposition, the auxiliary qubit remains in the same state after the gate application. Therefore, if we ensure the auxiliary qubit is in the state $\ket{0}$ before applying the quantum gates, we can reduce the decomposition to at most $12n$ CNOTs (\ref{C3}). This aligns with the optimization proposed in Section~\ref{subsec:fixing_phase}, which requires an auxiliary qubit in the state $\ket{0}$ and maintains the auxiliary qubit state, allowing the same auxiliary qubit to be reused for all decompositions. The intuition behind this optimization is that the first controlled gate flips the auxiliary qubit from $\ket{0}$ to $\ket{1}$ only if the first half of the control qubits are in the state $\ket{1}$. Consequently, the target qubit flips only if the second half of the control qubits and the auxiliary qubit are in the state $\ket{1}$. The third controlled operation flips the auxiliary qubit back to $\ket{0}$, preventing the execution of the last controlled operation.

To decompose the multi-controlled Pauli~$Y$ and $Z$ gates, given that we have an auxiliary qubit, we can use the multi-controlled Pauli~$X$ gate decomposition, adding unitary operations around the target qubit. For instance, to implement a multi-controlled $Z$ gate, we can apply a Hadamard gate around the target, since $HXH = Z$. Similarly, for a multi-controlled $Y$ gate, we can use a Phase gate with $HP(\tfrac{\pi}{2})XP(\tfrac{-\pi}{2})H = Y$. Therefore, any $n$-controlled Pauli gate can be decomposed with at most $12n$ CNOTs.

\begin{figure}
    \begin{equation*}
        \Qcircuit @C=1em @R=1em {
        & \qw {/} & \ctrl{2}            & \qw &   & & \qw {/}^{k_0} & \qw              & \ctrl{2} & \qw      & \qw      & \qw              & \ctrl{2} & \qw      & \qw     & \qw              & \qw              & \qw \\
        & \qw {/} & \ctrl{1}            & \qw & = & & \qw {/}^{k_1} & \qw              & \qw      & \qw      & \ctrl{1} & \qw              & \qw      & \qw      & \ctrl{1}& \qw              & \qw              & \qw \\
        & \qw     & \gate{\overline{U}} & \qw &   & & \qw           & \gate{V^\dagger} & \targ    & \gate{A} & \targ    & \gate{A^\dagger} & \targ    & \gate{A} & \targ   & \gate{A^\dagger} & \gate{V} & \qw \\
        }
    \end{equation*}
    \caption{Multi-controlled $SU(2)$ gate decomposition with $n$-control qubits, where $k_0 = \lfloor{\tfrac{n}{2}}\rfloor$, $k_1 = n - k_0$, $\overline{U} = VDV^\dagger$, and $D = A^\dagger XA X A^\dagger X A X$.}
    \label{fig:su2_decomposition}
\end{figure}
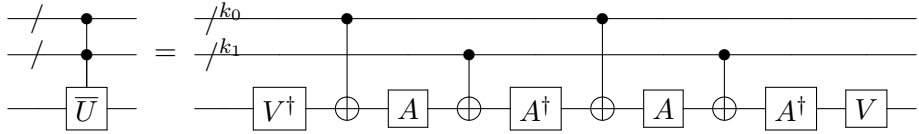

The decomposition of multi-controlled $SU(2)$ gates builds on top of the presented multi-controlled Pauli~$X$ decomposition. For a gate $\overline{U} \in SU(2)$ with at least one real diagonal element, \citet{valeCircuitDecompositionMulticontrolled2024} proposes a decomposition that takes at most $16n$ CNOTs, and up to $20n$ CNOTs when generalized for any $SU(2)$ gate. This generalization uses the eigendecomposition $\overline{U} = VDV^\dagger$, where the final controlled gate is $C^n\overline{U} = (C^nV)\cdot(C^nD)\cdot(C^nV^\dagger)$. The decomposition of $(C^nD)$ can follow the same schema as the decomposition of a multi-controlled $SU(2)$ gate with a real diagonal, but due to CNOT cancellations that can occur in $(C^nD)\cdot(C^nV^\dagger)$, the entire decomposition takes at most $20n$ CNOTs. However, leveraging the proposed optimization~\ref{C4}, there is no need to apply the $V$ unitary with control qubits, since if the controlled-$D$ is not applied, it results in the identity operation because $VV^\dagger = I$. Therefore, there is no need to decompose the $(C^nV^\dagger)$ and $(C^nV)$ gates, and the final decomposition takes at most $16n$ CNOTs (\ref{C2}). Figure~\ref{fig:su2_decomposition} presents the schema for decomposing a multi-controlled $SU(2)$ gate. Note that the eigendecomposition is not necessary when the gate has a real diagonal, although it streamlines the calculation of the $A$ gate since it results in an $R_Z$ gate, which can be performed virtually in superconducting qubits~\cite{mckayEfficientGatesQuantum2017}.

The decomposition proposed by \citet{dasilvaLineardepthQuantumCircuits2022} works for any multi-controlled single-target quantum gate, and the resulting circuit has linear depth, meaning that, by executing the gates in parallel when possible, it takes linear time in the quantum computer. Additionally, this decomposition has the benefit of not requiring any auxiliary qubits. However, it adds a quadratic number of CNOTs, resulting in a time complexity of $O(n^2)$ to execute this algorithm on a classical computer. Furthermore, this schema requires computing fractions from $\pi$ to $\tfrac{\pi}{2^{n-1}}$, which accumulates floating-point errors. In the following section, we will cover how to rewrite a multi-controlled $U(2)$ gate as a sequence of multi-controlled $SU(2)$ gates, allowing us to use the $O(n)$ decomposition by adding a single auxiliary qubit.

\section{Quantum Gate Rewrite}
\label{sec:replace}

In this section, we explore how to rewrite any $U(2)$ gate in terms of $SU(2)$ gates to leverage the efficient multi-controlled $SU(2)$ gate decomposition for any multi-controlled $U(2)$ gate. For any $2 \times 2$ unitary matrix $U$, there exists a corresponding $2 \times 2$ special unitary matrix $\overline{U}$, such that\footnote{We conclude that Equation~\eqref{eq:uu2} is valid based on \citet[Theorem 4.1]{nielsenQuantumComputationQuantum2010}, since the product of $SU(2)$ gates, in the case of rotation gates, is also an element of $SU(2)$.}  
\begin{equation}
    U = e^{i\phi} \overline{U}. \label{eq:uu2}
\end{equation}
To determine the phase angle $\phi$, we recall that since $U$ is unitary, its determinant satisfies $\det(U) = e^{i\Phi}$. Moreover, since $\overline{U}$ is a special unitary matrix, we have $\det(\overline{U}) = 1$. Substituting these properties into the determinant equation, we obtain  
\begin{equation}
    \begin{aligned}
        \det(U) &= \det(e^{i\phi} \overline{U}) \\
        e^{i\Phi} &= e^{i2\phi} \det(\overline{U}) = e^{i2\phi}.
    \end{aligned}
\end{equation}
Thus, it follows that $2\phi = \Phi$. The angle $\Phi$ can be determined as the phase of the complex number $e^{i\Phi}$, given by
\begin{equation}
    \Phi = \arctan\!2\big(\text{Im}(e^{i\Phi}),\ \text{Re}(e^{i\Phi})\big).
\end{equation}
Finally, solving for $\phi$, we obtain  
\begin{equation}
    \phi = \frac{1}{2} \Phi = \frac{1}{2} \arctan\!2\big(\text{Im}(\det U),\ \text{Re}(\det U)\big).
\end{equation}

Therefore, we can rewrite the $U(2)$ gates from Table~\ref{tab:quantum_gates} using the $SU(2)$ rotation gates. The results of these conversions are presented in Table~\ref{tab:su2_equivalent_gates}. While the global phase difference between $U$ and $\overline{U}$ has no measurement effect and can be ignored, for controlled operations, this global phase results in a relative phase difference. For instance, while $U= e^{i\phi}\overline{U}\equiv\overline{U}$,
\begin{align}
    C^1U = \begin{bNiceMatrix}
               I & \mathbf{0} \\ \mathbf{0} & U
           \end{bNiceMatrix} & \not \equiv
    e^{i\phi}C^1\overline{U} = \begin{bNiceMatrix}
                                     e^{i\phi}I & \mathbf{0} \\ \mathbf{0} & U
                                 \end{bNiceMatrix} \ \text{and} \\
    C^1U = \begin{bNiceMatrix}
               I & \mathbf{0} \\ \mathbf{0} & U
           \end{bNiceMatrix} & \not \equiv
    C^1\overline{U} = \begin{bNiceMatrix}
                          I & \mathbf{0} \\ \mathbf{0} & \overline{U}
                      \end{bNiceMatrix}.
\end{align}

\begin{table}[htbp]
    \caption{$U(2)$ and $SU(2)$ quantum gate equivalence.}
    \label{tab:su2_equivalent_gates}
    \centering
    \small
    \begin{tabular}{lcc}
        \toprule
        \textbf{$U(2)$ Gate} & \textbf{$SU(2)$ Equivalent}            & \textbf{Global Phase}    \\
        \midrule
        Pauli~$X$            & $X \equiv R_X(\pi)$                    & $e^{i\tfrac{\pi}{2}}$    \\
        Pauli~$Y$            & $Y \equiv R_Y(\pi)$                    & $e^{i\tfrac{\pi}{2}}$    \\
        Pauli~$Z$            & $Z \equiv R_Z(\pi)$                    & $e^{i\tfrac{\pi}{2}}$    \\
        Phase                & $P(\theta) \equiv R_Z(\theta)$         & $e^{i\tfrac{\theta}{2}}$ \\
        Hadamard             & $H \equiv R_X(\pi)R_Y(\tfrac{\pi}{2})$ & $e^{i\tfrac{\pi}{2}}$    \\
        \hline
    \end{tabular}
\end{table}

As presented, the global phase difference between a $U(2)$ and $SU(2)$ gate matters when using the gate in a controlled operation. Therefore, the simple rewrite of $U(2)$ into an $SU(2)$ is not sufficient to leverage the $SU(2)$ more efficient decomposition. A contribution of this paper is how to fix this phase difference and use the linear time decomposition algorithms for any quantum gate. We develop our solution in the next subsection. First, in Subsection~\ref{subsec:fixing_phase}, we provide a simple example of how we can fix the phase of a single-controlled gate and generalize the result for multi-controlled gates. Although the phase fixing results in the correct quantum gate, this workaround negates the proposed rewrite optimization. Therefore, in Subsection~\ref{subsec:aux_qubit}, we introduce an auxiliary qubit in the decomposition to enable the linear decomposition of any multi-controlled $U(2)$ gate. This optimization is especially beneficial for decomposition multi-controlled Phase and Hadamard gates.

\subsection{Fixing the Phase}
\label{subsec:fixing_phase}

For a single-controlled quantum gate, considering that $U = e^{i\phi}\overline{U}$, we can see that a $P(\phi)$ gate applied to the control qubits can fix the phase.

\begin{equation}
    \begin{aligned}
        C^1U & = \left(P(\phi)\otimes I\right) \cdot C^1\overline{U}                                                                   \\
             & = \left(P(\phi)\otimes I\right) \cdot \left({\ket{0}}{\bra{0}}\otimes I + {\ket{1}}{\bra{1}}\otimes \overline{U}\right) \\
             & = P(\phi){\ket{0}}{\bra{0}}\otimes I + P(\phi){\ket{1}}{\bra{1}}\otimes \overline{U}                                  \\
             & = {\ket{0}}{\bra{0}}\otimes I + e^{i\phi}{\ket{1}}{\bra{1}}\otimes \overline{U}                                         \\
             & = {\ket{0}}{\bra{0}}\otimes I + {\ket{1}}{\bra{1}}\otimes U.
    \end{aligned}
\end{equation}

The intuition for this equation is that the phase must be applied only when the gate $\overline{U}$ is applied. In the controlled version, this is when the control qubit is in the state $\ket{1}$. Since the Phase gate has no effect on the $\ket{0}$ state, the phase gate applies a phase $e^{i\phi}$ when the control is in the state $\ket{1}$, corresponding to the application of the gate $\overline{U}$. Note that despite the phase being applied to the control qubits instead of the target, it has the same result as $(\alpha\cdot\ket{\psi})\otimes\ket{\varphi}= \ket{\psi}\otimes(\alpha\cdot\ket{\varphi})$.

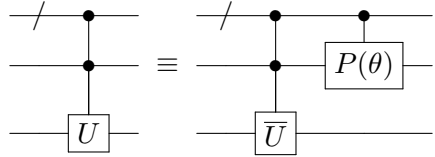
\begin{figure}[tp]
    \begin{equation*}
        \Qcircuit @C=1em @R=.8em {
        & {/} \qw       & \ctrl{2} & \qw &         &    &  {/} \qw & \ctrl{2} &  \ctrl{1}         & \qw\\
        & \qw           & \ctrl{1} & \qw & \equiv  &    & \qw      & \ctrl{1} &  \gate{P(\phi)} & \qw \\
        & \qw           & \gate{U} & \qw &         &    & \qw      & \gate{\overline{U}} &  \qw              & \qw
        }
    \end{equation*}
    \caption{Multi-controlled $U(2)$ gate rewrite as a multi-controlled $SU(2)$ gate using a $C^{n-1}P(\phi)$ gate for phase correction.}
    \label{fig:cn-1p}
\end{figure}

Generalizing for $n$-control qubits, we can see that the circuit illustrated in Figure~\ref{fig:cn-1p} using $C^{n-1}P(\phi)$ gate can fix the phase as
\begin{equation}
    \begin{aligned}
        C^nU  = & \left(C^{n-1}P(\phi)\otimes I\right)\cdot C^n\overline{U}                                                                                                                          \\
        =       & \left(\overbrace{ \begin{bNiceMatrix}[]
                                            1      & 0      & \Cdots & 0      & 0           \\
                                            0      & 1      & \Cdots & 0      & 0           \\
                                            \Vdots & \Vdots & \Ddots & \Vdots & \Vdots      \\
                                            0      & 0      & \Cdots & 1      & 0           \\
                                            0      & 0      & \Cdots & 0      & e^{i\phi} \\
                                        \end{bNiceMatrix}}^{2^{n}\times2^{n}} \otimes \begin{bNiceMatrix}
                                                                                      1 & 0 \\
                                                                                      0 & 1
                                                                                  \end{bNiceMatrix}\right)\cdot \overbrace{\begin{bNiceMatrix}[]
                                                                                                                                   1      & 0      & \Cdots & 0      & 0                  & 0                  \\
                                                                                                                                   0      & 1      & \Cdots & 0      & 0                  & 0                  \\
                                                                                                                                   \Vdots & \Vdots & \Ddots & \Vdots & \Vdots             & \Vdots             \\
                                                                                                                                   0      & 0      & \Cdots & 1      & 0                  & 0                  \\
                                                                                                                                   0      & 0      & \Cdots & 0      & \overline{U}_{0,0} & \overline{U}_{0,1} \\
                                                                                                                                   0      & 0      & \Cdots & 0      & \overline{U}_{1,0} & \overline{U}_{1,1} \\
                                                                                                                               \end{bNiceMatrix}}^{2^{n+1}\times2^{n+1}} \\
        =       & \underbrace{\begin{bNiceMatrix}[small]
                                      1      & 0      & \Cdots & 0      & 0           & 0           \\
                                      0      & 1      & \Cdots & 0      & 0           & 0           \\
                                      \Vdots & \Vdots & \Ddots & \Vdots & \Vdots      & \Vdots      \\
                                      0      & 0      & \Cdots & 1      & 0           & 0           \\
                                      0      & 0      & \Cdots & 0      & e^{i\phi} & 0           \\
                                      0      & 0      & \Cdots & 0      & 0           & e^{i\phi} \\
                                  \end{bNiceMatrix}}_{2^{n+1}\times2^{n+1}}\cdot \begin{bNiceMatrix}[small]
                                                                                 1      & 0      & \Cdots & 0      & 0                  & 0                  \\
                                                                                 0      & 1      & \Cdots & 0      & 0                  & 0                  \\
                                                                                 \Vdots & \Vdots & \Ddots & \Vdots & \Vdots             & \Vdots             \\
                                                                                 0      & 0      & \Cdots & 1      & 0                  & 0                  \\
                                                                                 0      & 0      & \Cdots & 0      & \overline{U}_{0,0} & \overline{U}_{0,1} \\
                                                                                 0      & 0      & \Cdots & 0      & \overline{U}_{1,0} & \overline{U}_{1,1} \\
                                                                             \end{bNiceMatrix}
        =        \begin{bNiceMatrix}[]
                     1      & 0      & \Cdots & 0      & 0       & 0       \\
                     0      & 1      & \Cdots & 0      & 0       & 0       \\
                     \Vdots & \Vdots & \Ddots & \Vdots & \Vdots  & \Vdots  \\
                     0      & 0      & \Cdots & 1      & 0       & 0       \\
                     0      & 0      & \Cdots & 0      & U_{0,0} & U_{0,1} \\
                     0      & 0      & \Cdots & 0      & U_{1,0} & U_{1,1} \\
                 \end{bNiceMatrix}.
    \end{aligned}
\end{equation}

We can follow the same intuition as in the single-control case, with the phase being applied only when all the control qubits are in the state $\ket{1}$. Note that for the multi-controlled case, the $C^{n-1}P(\phi)$ must be decomposed into one and two-qubit gates to execute on the quantum computer. However, since the Phase gate is not an $SU(2)$ gate, this decomposition takes quadratic compilation time and space, which eliminates the benefit of rewriting the $U$ gate as $SU(2)$. In the next subsection, we introduce an auxiliary qubit and substitute the multi-control Phase gate with a multi-control $Z$ rotation gate, therefore making all decompositions linear.

\subsection{Adding an Auxiliary Qubit}
\label{subsec:aux_qubit}

With an auxiliary qubit initialized in the state $\ket{0}$, we can use a controlled $R_Z$ gate to introduce a phase $e^{i\phi}$ after applying a $C^n\overline{U}$ gate, thereby implementing a $C^nU$ gate. Starting with a single control, we can see that appending a $C^1_cR_Z(-2\phi)_a$ after the controlled-$\overline{U}$ gate fixes the phase. To denote the control and target qubits of an operation, we use subscripts, with $C^1_c\overline{U}_t$ meaning the qubit $c$ controls the application of the gate $\overline{U}$ on the qubit $t$, and $C^1_cR_Z(-2\phi)_a$ meaning the qubit $c$ controls the gate $R_Z$ on qubit $a$, where $c$, $t$, and $a$ stand for control, target, and auxiliary, respectively. Expanding the rewriting, we see that the application is equivalent to

\begin{figure}[htbp]
    \begin{equation*}
        \Qcircuit @C=1em @R=.8em {
        \lstick{\ket{c}} & {/} \qw & \ctrl{1} & \qw &        & & & \lstick{\ket{c}}  & {/} \qw & \ctrl{1}            & \ctrl{2}             & \qw \\
        \lstick{\ket{t}} &     \qw & \gate{U} & \qw & \equiv & & & \lstick{\ket{t}}  &     \qw & \gate{\overline{U}} & \qw                  & \qw \\
        &         &          &     &                         & & &   &            & \lstick{\ket{a} = \ket{0}} & \gate{R_Z(-2\phi)} & \qw
        }
    \end{equation*}

    \caption{Multi-controlled single-target $U(2)$ gate rewrite to take advantage of $SU(2)$ decomposition.}    \label{fig:cnrx}
\end{figure}
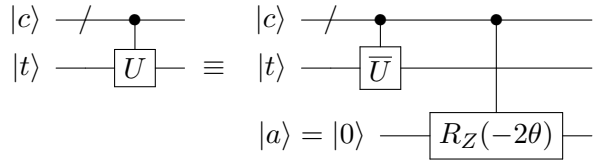

\begin{equation}
    \begin{aligned}
        C^1_c U_t  \equiv & C^1_cR_Z(-2\phi)_a \cdot C^1_c\overline{U}_t                                                                                                                                                                                                               \\
        =                 & \begin{aligned}[t] &
               \left({\ket{0}}{\bra{0}}_c\otimes I_t \otimes I_a + {\ket{1}}{\bra{1}}_c\otimes I_t \otimes R_Z(-2\phi)_a \right)              \\
                   & \cdot \left({\ket{0}}{\bra{0}}_c\otimes I_t \otimes I_a + {\ket{1}}{\bra{1}}_c  \otimes \overline{U}_t \otimes I_a \right)
                            \end{aligned} \\
        =                 & {\ket{0}}{\bra{0}}_c\otimes I_t \otimes I_a + {\ket{1}}{\bra{1}}_c  \otimes \overline{U}_t \otimes R_Z(-2\phi)_a                                                                                                                                           \\
        =                 & {\ket{0}}{\bra{0}}_c\otimes I_t \otimes I_a + {\ket{1}}{\bra{1}}_c  \otimes \overline{U}_t \otimes  \left(e^{i\phi}{\ket{0}}{\bra{0}}_a + e^{-i\phi}{\ket{1}}{\bra{1}}_a\right).
    \end{aligned}
\end{equation}
Since the auxiliary qubit is in the state $\ket{0}$, projecting it to $\ket{0}$ does not change the quantum state but allows us to state that
\begin{equation}
    \begin{aligned}
        C^1_cU_t \equiv & \left({C^1}_cR_Z(-2\phi)_a {C^1}_c\overline{U}_t\right) \cdot \left(I \otimes I \otimes {\ket{0}}{\bra{0}}\right)                                  \\
        =               & {\ket{0}}{\bra{0}}_c\otimes I_t \otimes {\ket{0}}{\bra{0}}_a + {\ket{1}}{\bra{1}}_c  \otimes \overline{U}_t \otimes  e^{i\phi}{\ket{0}}{\bra{0}}_a \\
        =               & ({\ket{0}}{\bra{0}}_c\otimes I_t + {\ket{1}}{\bra{1}}_c  \otimes U_t) \otimes  {\ket{0}}{\bra{0}}_a.
    \end{aligned}
\end{equation}

The intuition for this rewrite builds on the previous subsection. However, since the $R_Z$ gate changes the phase of both $\ket{0}$ and $\ket{1}$ states, we need an auxiliary qubit that we know to be in the state $\ket{0}$, thereby selecting the desired phase from the $R_Z$ gate. Note that the auxiliary qubit remains in the $\ket{0}$ state, allowing us to reuse it in other operations.

We can straightforwardly extend this concept to multiple control qubits, resulting in the unitary matrix\footnote{In Equation~\eqref{eq:czcu}, we use integer notation inside the kets for the control qubits, where $\ket{l}$ represents an $n$-qubit state.}
\begin{equation}
    \begin{aligned}
        C^n_cU_t  \equiv & C^n_cR_Z(-2\phi)_a \cdot C^n_c\overline{U}_t                                                                                                                                                                                                                             \\
        =                & \begin{aligned}[t]
                                & \left(\sum_{l=0}^{2^n-2}{\ket{l}}{\bra{l}}_c\otimes I_t \otimes I_a +{\ket{2^n-1}}{\bra{2^n-1}}_c\otimes I_t \otimes R_Z(-2\phi)_a \right)        \\
                                & \cdot \left(\sum_{k=0}^{2^n-2}{\ket{k}}{\bra{k}}_c\otimes I_t \otimes I_a +{\ket{2^n-1}}{\bra{2^n-1}}_c  \otimes \overline{U}_t \otimes I_a \right)
                           \end{aligned} \\
        =                & \overbrace{\begin{bNiceMatrix}[]
                                              1      & \Cdots & 0      & 0           & 0            & 0           & 0            \\
                                              \Vdots & \Ddots & \Vdots & \Vdots      & \Vdots       & \Vdots      & \Vdots       \\
                                              0      & \Cdots & 1      & 0           & 0            & 0           & 0            \\
                                              0      & \Cdots & 0      & e^{i\phi} & 0            & 0           & 0            \\
                                              0      & \Cdots & 0      & 0           & e^{-i\phi} & 0           & 0            \\
                                              0      & \Cdots & 0      & 0           & 0            & e^{i\phi} & 0            \\
                                              0      & \Cdots & 0      & 0           & 0            & 0           & e^{-i\phi} \\
                                          \end{bNiceMatrix}}^{2^{n+2}\times2^{n+2}}
        \cdot\begin{bNiceMatrix}[]
                 1      & \Cdots & 0      & 0                  & 0                  & 0                  & 0                  \\
                 \Vdots & \Ddots & \Vdots & \Vdots             & \Vdots             & \Vdots             & \Vdots             \\
                 0      & \Cdots & 1      & 0                  & 0                  & 0                  & 0                  \\
                 0      & \Cdots & 0      & \overline{U}_{0,0} & 0                  & \overline{U}_{0,1} & 0                  \\
                 0      & \Cdots & 0      & 0                  & \overline{U}_{0,0} & 0                  & \overline{U}_{0,1} \\
                 0      & \Cdots & 0      & \overline{U}_{1,0} & 0                  & \overline{U}_{1,1} & 0                  \\
                 0      & \Cdots & 0      & 0                  & \overline{U}_{1,0} & 0                  & \overline{U}_{1,1} \\
             \end{bNiceMatrix}                                                                                                                                                                             \\
        =                & \begin{bNiceMatrix}[]
                               1      & \Cdots & 0      & 0       & 0                              & 0       & 0                              \\
                               \Vdots & \Ddots & \Vdots & \Vdots  & \Vdots                         & \Vdots  & \Vdots                         \\
                               0      & \Cdots & 1      & 0       & 0                              & 0       & 0                              \\
                               0      & \Cdots & 0      & U_{0,0} & 0                              & U_{0,1} & 0                              \\
                               0      & \Cdots & 0      & 0       & e^{-i\phi}\overline{U}_{0,0} & 0       & e^{-i\phi}\overline{U}_{0,1} \\
                               0      & \Cdots & 0      & U_{1,0} & 0                              & U_{1,1} & 0                              \\
                               0      & \Cdots & 0      & 0       & e^{-i\phi}\overline{U}_{1,0} & 0       & e^{-i\phi}\overline{U}_{1,1} \\
                           \end{bNiceMatrix}.
    \end{aligned}
    \label{eq:czcu}
\end{equation}

Since we know that the auxiliary qubit is in the state $\ket{0}$, we can project into $\ket{0}$ and trace out the auxiliary qubit, resulting in exactly the matrix for a multi-controlled $U$ gate:

\begin{equation}
    \begin{aligned}
        {C^n}_cU_t  \equiv & \left({C^n}_cR_Z(-2\phi)_a \cdot {C^n}_c\overline{U}_t \right)       \cdot \left({I^{\otimes n}}_c \otimes I_t \otimes{\ket{0}}{\bra{0}}_a\right) \\
        =                  & \begin{bNiceMatrix}[]
                                 1      & 0      & 0      & \Cdots & 0      & 0       & 0      & 0       & 0      \\
                                 0      & 0      & 0      & \Cdots & 0      & 0       & 0      & 0       & 0      \\
                                 0      & 0      & 1      & \Cdots & 0      & 0       & 0      & 0       & 0      \\
                                 \Vdots & \Vdots & \Vdots & \Ddots & \Vdots & \Vdots  & \Vdots & \Vdots  & \Vdots \\
                                 0      & 0      & 0      & \Cdots & 0      & 0       & 0      & 0       & 0      \\
                                 0      & 0      & 0      & \Cdots & 0      & U_{0,0} & 0      & U_{0,1} & 0      \\
                                 0      & 0      & 0      & \Cdots & 0      & 0       & 0      & 0       & 0      \\
                                 0      & 0      & 0      & \Cdots & 0      & U_{1,0} & 0      & U_{1,1} & 0      \\
                                 0      & 0      & 0      & \Cdots & 0      & 0       & 0      & 0       & 0      \\
                             \end{bNiceMatrix}
        \xRightarrow{\text{Tr}_a}  \overbrace{\begin{bNiceMatrix}[]
                                                      1      & 0      & \Cdots & 0      & 0       & 0       \\
                                                      0      & 1      & \Cdots & 0      & 0       & 0       \\
                                                      \Vdots & \Vdots & \Ddots & \Vdots & \Vdots  & \Vdots  \\
                                                      0      & 0      & \Cdots & 1      & 0       & 0       \\
                                                      0      & 0      & \Cdots & 0      & U_{0,0} & U_{0,1} \\
                                                      0      & 0      & \Cdots & 0      & U_{1,0} & U_{1,1} \\
                                                  \end{bNiceMatrix}}^{2^{n+1}\times2^{n+1}} = {C^n}_cU_t
    \end{aligned}
\end{equation}

Figure~\ref{fig:cnrx} illustrates the proposed rewrite. Note that, since the $\overline{U}$ and $R_Z$ gates are in $SU(2)$, the decomposition of the resulting circuit has a lower time complexity (with $16n$ CNOTs for each one) than the original circuit (with $O(n^2)$ CNOTs), requiring at most $32n$ CNOTs (\ref{C1}). Also, since the auxiliary qubit can be reused, the overhead of this rewriting is only a single auxiliary qubit for the entire quantum circuit. This result improves upon previous linear time decompositions. For instance, the proposal by \citet[Lemma 7.11]{barencoElementaryGatesQuantum1995} requires two auxiliary qubits, while the decomposition presented by \citet[p. 184]{nielsenQuantumComputationQuantum2010} requires exactly $n-1$ auxiliary qubits.

For a multi-controlled Phase gate, considering that $P(\theta) = e^{i\frac{\theta}{2}}R_Z(\theta)$, we can rewrite this operation with a single multi-controlled $R_Z$ gate and one auxiliary qubit, as illustrated in Figure~\ref{fig:cnp}. By using the target qubit of the original instruction also as a control in the multi-controlled $R_Z$, the auxiliary qubit ends up in the state $R_Z(-2\theta)\ket{0} = e^{i\theta}\ket{0}$. Since the $P(\theta)$ gate only applies the relative phase $e^{i\theta}$ to the $\ket{1}$ state, particularly the control state, the target qubit acquires the phase $e^{i\theta}$ only if all the control qubits and the target qubit are in the state $\ket{1}$. With this scheme, the decomposition of a multi-controlled Phase gate takes at most $16n$ CNOTs.

\begin{figure}[h]

    \begin{equation*}
        \Qcircuit @C=1em @R=.8em {
        \lstick{\ket{c}} & {/} \qw & \ctrl{1}         & \qw &        & & & \lstick{\ket{c}}         & {/} \qw & \qw                        & \ctrl{2}             & \qw \\
        \lstick{\ket{t}} &     \qw & \gate{P(\theta)} & \qw & \equiv & & & \lstick{\ket{t}}         &     \qw & \qw                        & \ctrl{1}             & \qw \\
        &         &                  &     &        & & &                          &         & \lstick{\ket{a} = \ket{0}} & \gate{R_Z(-2\theta)} & \qw \\
        }
    \end{equation*}
    \caption{Multi-controlled $P(\theta)$ gate rewrite to take advantage of $SU(2)$ decomposition.}    \label{fig:cnp}

\end{figure}

In the next section, we discuss the impact of this and other optimizations proposed in this paper on the decomposed quantum circuit. Additionally, we evaluate the implementation of these optimizations in the Ket quantum programming platform.

\section{Results}
\label{sec:results}

Summarizing the optimizations proposed in this article, Table~\ref{tab:cnots} shows that given an auxiliary qubit in the state $\ket{0}$, we can decompose any $n$-controlled quantum gate from Table~\ref{tab:quantum_gates} using a linear number of quantum gates CNOTs. The gates that benefit most from this optimization are the multi-controlled Phase and Hadamard gates, whose time complexity has been reduced from quadratic to linear given one auxiliary qubits. This reduction not only impacts the quantum runtime but also the classical runtime, as the classical computer computes the decompositions. For the rotation gates, since these $SU(2)$ gates fall under the special case presented in \citet{valeCircuitDecompositionMulticontrolled2024}, having at least one real diagonal, the time complexity for decomposing their multi-controlled version remains the same.

\begin{table}[htbp]
    \centering
    \caption{Comparison of the number of CNOTs required for $n$-controlled single-qubit gate decomposition. The subscript in the number of auxiliary qubits indicates whether the qubit can be dirty ($d$) or must be in the $\ket{0}$ state ($0$). Our results are marked in bold.}
    \label{tab:cnots}
    \small
    \begin{tabular}{lcccccc}
        \toprule
        \multirow{3}{*}{\textbf{Gate Name}} & \multicolumn{6}{c}{\textbf{Number of CNOTs/Auxiliary Qubits}}                                                                                                                                                                                                                                                    \\
        \cmidrule{2-7}
                           &                                                               & \multicolumn{5}{c}{\textbf{Comparison with State-of-the-Art}}                                                                                                                                                                                    \\
        \cmidrule{3-7}
                           & \textbf{Our}                                                  & \cite{valeCircuitDecompositionMulticontrolled2024}            & \cite{itenQuantumCircuitsIsometries2016} & \cite{dasilvaLineardepthQuantumCircuits2022} & \cite{barencoElementaryGatesQuantum1995} & \cite{nielsenQuantumComputationQuantum2010} \\
        \midrule
        Pauli~$X$          & $\boldsymbol{12n/1_0}$                                        &                                                               & $16n/1_d$; $\ 4n/n-2_d$                  & $O(n^2)/0$                                   &                                                                                        \\
        Pauli~$Y$          & $\boldsymbol{12n/1_0}$                                        &                                                               & $16n/1_d$; $\ 4n/n-2_d$                  & $O(n^2)/0$                                   &                                                                                        \\
        Pauli~$Z$          & $\boldsymbol{12n/1_0}$                                        &                                                               & $16n/1_d$; $\ 4n/n-2_d$                  & $O(n^2)/0$                                   &                                                                                        \\
        Hadamard           & $\boldsymbol{32n/1_0}$                                        &                                                               &                                          & $O(n^2)/0$                                   & $32n/2_{0,d}$                            & $2n/n_0$                                    \\
        Phase              & $\boldsymbol{16n/1_0}$                                        &                                                               &                                          & $O(n^2)/0$                                   & $32n/2_{0,d}$                            & $2n/n_0$                                    \\
        $X$ Rotation       &                                                               & $16n/0$                                                                                                                                                                                                                                          \\
        $Y$ Rotation       &                                                               & $16n/0$                                                                                                                                                                                                                                          \\
        $Z$ Rotation       &                                                               & $16n/0$                                                                                                                                                                                                                                          \\
        \bottomrule
    \end{tabular}
\end{table}

In this section, we will explore how these optimizations impact the CNOT count of two algorithms: Grover's algorithm, which requires multi-controlled Pauli~$Z$ gates, and a state preparation algorithm that requires multi-controlled Phase gates. For our evaluation, we will compare the final number of CNOTs with and without the auxiliary qubit. All evaluations were performed using the Ket quantum programming platform, where we implemented all the decompositions and optimizations presented in this paper (\ref{C5}).

Note that enabling these optimizations requires no changes to the code. In Ket's architecture, the quantum execution target determines which features will be available for execution, including whether decomposition is necessary. By default, the \texttt{Process} class uses simulation, which does not require decomposition~\cite{darosaQSystemBitwiseRepresentation2020}. 

We have prepared a repository\footnote{\url{https://gitlab.com/quantum-ket/arxiv-2406.05581}} containing the Ket version and the data generated in our experiments. Note that the Ket version used in this study does not match any official platform release. In this version, a \texttt{decompose} parameter is provided in the \texttt{Process} class constructor for testing purposes, enabling the optimization by default when activated. Additionally, when decomposition is enabled and the optimization is not explicitly disabled\footnote{To disable the optimization, set the environment variable \texttt{\tiny KET\_DISABLE\_DECOMPOSITION\_OPTIMIZATION=true.}}, the simulation will execute with one additional qubit beyond what is specified in the constructor, allowing it to be used as an auxiliary qubit.

\subsection{Evaluation: Grover's Algorithm}

Grover's algorithm~\cite{groverFastQuantumMechanical1996} can find an element in an unstructured list in time $O(\sqrt{N})$, where $N$ is the number of elements in the list, using $n = \log_2(N)$ qubits. This algorithm has applications in various fields such as solving NP problems~\cite{cerfNestedQuantumSearch2000}, optimization~\cite{durrQuantumAlgorithmFinding1996}, and  cryptography~\cite{aaronsonQuantumLowerBounds2004,brassardQuantumCryptanalysisHash1998}. It works by marking the search state in superposition and increasing the probability of measuring it, doing so in $\sqrt{N}$ iterations.

\begin{figure}[htbp]
    \centering

    \hfill
    \begin{subfigure}[c]{.45\textwidth}
        \centering
        \begin{minipage}{.7\linewidth}
            \begin{minted}[frame=lines,fontsize=\tiny,linenos,breaklines]{py}
def oracle(qubits):
    """|11...1⟩ -> -|11...1⟩"""
    ctrl(qubits[:-1], Z)(qubits[-1])

def grover_layer(qubits):
    oracle(qubits)
    grover_diffusion(qubits)

def grover(size):
    process = Process(
        num_qubits=size,
        decompose=True, 
        execution="batch"
    )
    qubits = process.alloc(size)
    H(qubits)
    steps = int((pi / 4) * sqrt(2**size))
    for _ in range(steps):
        grover_layer(qubits)
    return measure(qubits)
    \end{minted}
        \end{minipage}
        \caption{Grover's Algorithm implemented using Ket. The Function \texttt{grover\_diffusion} implementation is presented in Figure~\ref{fig:ket_grover_op}.}
        \label{fig:ket_grover}
    \end{subfigure}
    \hfill
    \begin{subfigure}[c]{0.45\textwidth}
        \centering
        \includegraphics[width=\linewidth]{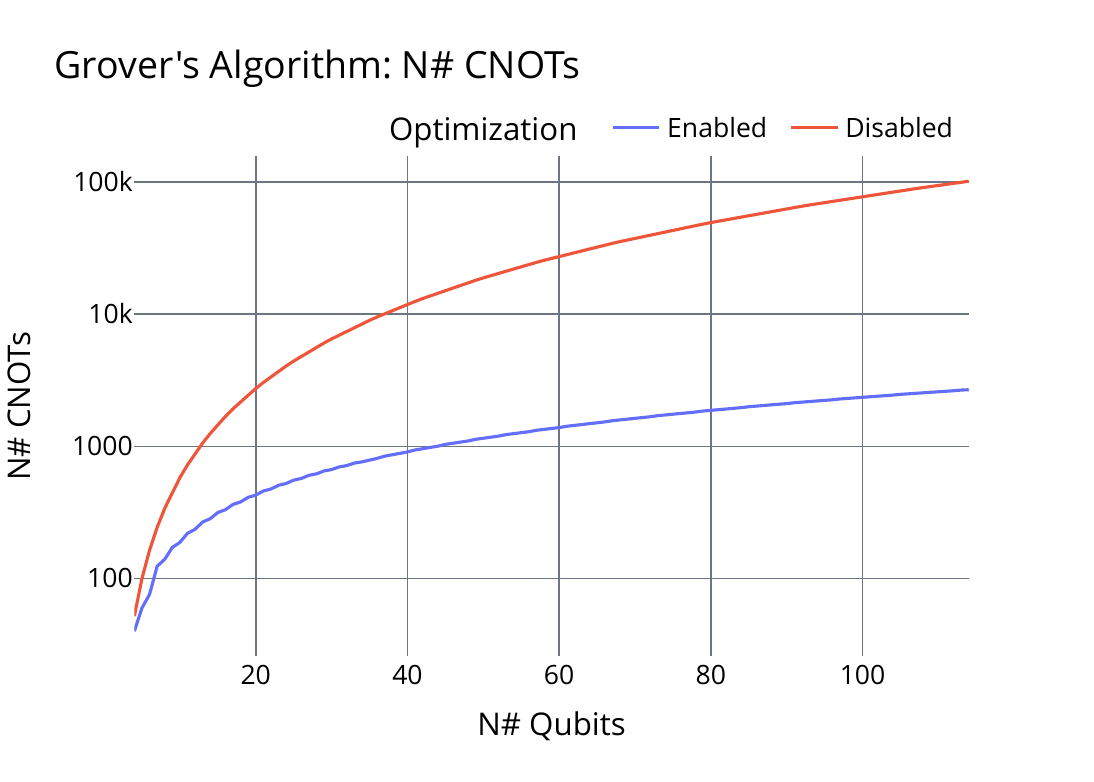}
        \caption{Number of CNOTs resulting from the decomposition of a single layer of Grover's algorithm.}
        \label{fig:grover_cnots}
    \end{subfigure}
    \hfill

    \caption{Grover's Algorithm evaluation.}
\end{figure}

Figure~\ref{fig:ket_grover} illustrates Grover's algorithm implemented using Ket. Each Grover layer contains two multi-controlled Pauli~$Z$ gates: one for the oracle and another for the diffusion operation. We used an oracle that marks the state $\ket{11\dots1}$; however, marking any other basis state only requires applying Pauli~$X$ gates around to certain qubits. To evaluate the decomposition, we executed a single layer of the algorithm, resulting in a decomposition with at most $24n$ CNOTs when optimization is enabled.  Even with just one layer, there is a substantial improvement in the number of CNOTs required for the decomposition.

Figure~\ref{fig:grover_cnots} illustrates the number of CNOT gates resulting from the decomposition when the algorithm is executed with 4 to 114 input qubits. Notably, enabling the optimization requires one additional auxiliary qubit. For instance, with 114 qubits, the optimization reduced the number of CNOTs from 101,252 to 2,684. Note that without the auxiliary qubit, only the rotation gates that belong to $SU(2)$ are decomposed with a linear decomposition, while the other gates are decomposed with the quadratic multi-controlled $U(2)$ decomposition.

\subsection{Evaluation: State Preparation Algorithm}

\begin{figure}[htbp]
    \centering

    \hfill
    \begin{subfigure}[b]{.45\textwidth}
        \centering
        \begin{minipage}{\linewidth}
            \begin{minted}[frame=lines,xleftmargin=24pt,fontsize=\tiny,linenos,breaklines]{py}
class ParamTree:
    def __init__(self, prob, amp):
        assert (len(prob)).bit_count() == 1
        total = sum(prob)
        assert total != 0
        prob = [p / total for p in prob]
        l_prob = prob[: len(prob) // 2]
        l_amp = amp[: len(prob) // 2]
        r_prob = prob[len(prob) // 2 :]
        r_amp = amp[len(prob) // 2 :]
        self.value = 2 * asin(sqrt(sum(r_prob)))
        if len(prob) > 2:
            self.l = ParamTree(l_prob, l_amp)
            self.r = ParamTree(r_prob, right_amp)
        else:
            self.l = None
            self.r = None
            self.phase0 = amp[0]
            self.phase1 = amp[1]

    def is_leaf(self):
        return self.l is None and self.r is None
    \end{minted}
        \end{minipage}
        \caption{Class that implement part of the logics necessary for the State Preparation Algorithm presented in Figure~\ref{fig:ket_prepare}.}
        \label{fig:ket_param_tree}
    \end{subfigure}
    \hfill
    \begin{subfigure}[b]{0.45\textwidth}
        \centering
        \includegraphics[width=\linewidth]{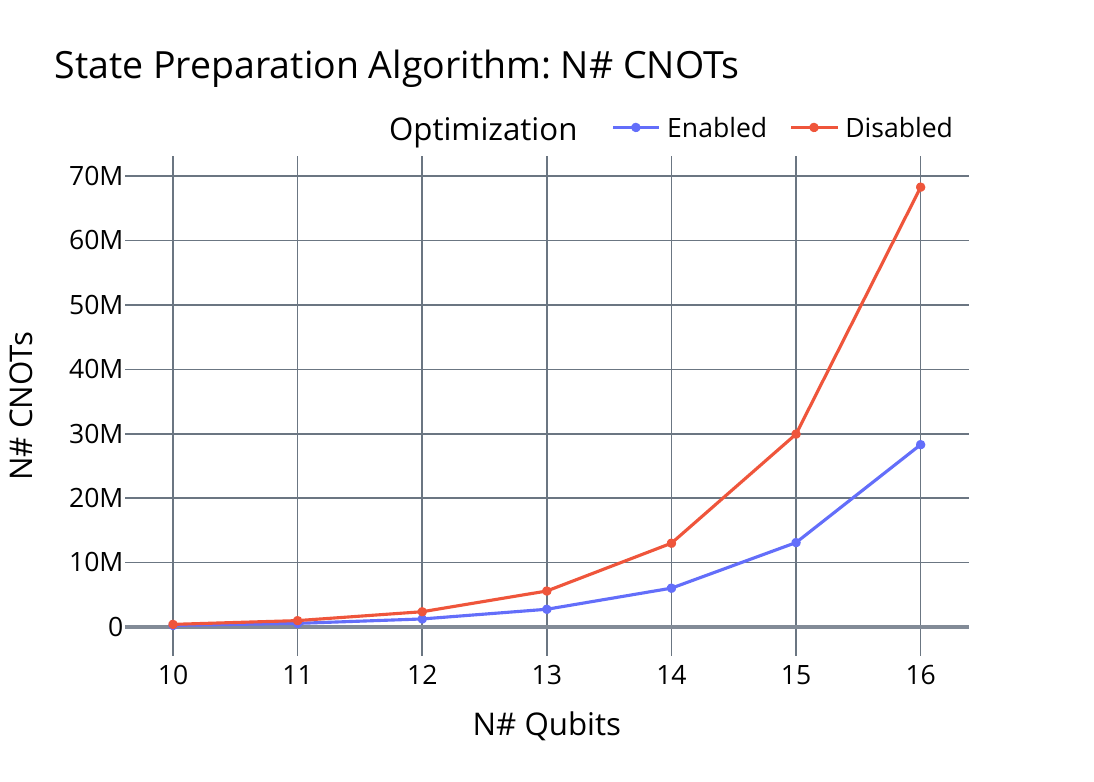}
        \caption{Number of CNOTs resulting from the decomposition of multi-controlled gates from the State Preparation Algorithm.}
        \label{fig:ket_param_cnots}
    \end{subfigure}
    \hfill
    \caption{Evaluation of the State Preparation Algorithm from Figure~\ref{fig:ket_prepare}.}
\end{figure}

Our next evaluation analyzes the impact of the proposed decomposition optimization in the State Preparation Algorithm~\cite{kitaevWavefunctionPreparationResampling2008} shown in Figure~\ref{fig:ket_prepare}. This algorithm prepares $n$-qubits in any possible quantum state, using two lists with $2^n$ parameters: one for the measurement probability and another for the phase of the probability amplitude. Although this algorithm requires an exponential number of parameters and its execution time grows exponentially with the number of qubits, making it unsuitable for solving practical problems, it is valuable for conducting quantum mechanics experiments with a quantum computer~\cite{zanettiSimulatingNoisyQuantum2023}. While this algorithm does not directly call for any controlled quantum gates, the recursion results in several multi-controlled $R_Y$ and Phase gates, as illustrated in Figure~\ref{fig:circuit_prepare}. Figure~\ref{fig:ket_param_cnots} shows the number of CNOTs resulting from the execution and decomposition of this algorithm, with 4 to 16 qubits. Notably, with the execution using 16 qubits, the optimization reduced the number of CNOTs from 68,288,004 to 28,311,044.

\section{Conclusion}
\label{sec:conclusion}

In this paper, we have demonstrated that any multi-controlled single-target quantum gate can be decomposed into a linear number of CNOT gates in linear time, with the expense of a single auxiliary qubit in the state $\ket{0}$. In the worst case, which includes the multi-controlled Hadamard gate, the decomposition takes at most $32n$ CNOT gates, while in the best case, such as for the multi-controlled Pauli gates, it takes at most $12n$ CNOTs. These efficient decompositions show that a quantum compiler can effectively decompose any quantum operation built with a single-qubit gate and the ability to add control qubits, as available in the Ket quantum programming platform. Therefore, decomposition should not be a concern for high-level quantum programming. However, quantum programmers will benefit from knowing that some quantum gates are more efficient than others, similar to how classical programmers know that a shift instruction takes fewer computing cycles in a CPU than a multiplication instruction.

One possible optimization for quantum circuits is combining the effect of consecutive multi-controlled operations that share the same control qubits and target, knowing that the upper bound for a decomposition is $32n$ CNOTs. This result comes from one of the main contributions of this paper: a method to rewrite $U(2)$ gates as $SU(2)$ gates with an auxiliary qubit to manage the phase differences in controlled operations. We have demonstrated that applying a controlled $R_Z$ gate on an auxiliary qubit, initialized in the $\ket{0}$ state, can effectively fix the phase discrepancy. This technique allows us to utilize the more efficient decomposition of $SU(2)$ for any $U(2)$ gates, maintaining the integrity of controlled operations.

In conclusion, the optimization proposed in this paper, particularly the method for rewriting $U(2)$ gates as $SU(2)$ gates with phase correction using an auxiliary qubit, offers a viable path to optimizing quantum circuit designs. By minimizing the gate count, especially CNOT gates, we can improve the performance and reliability of quantum algorithms. This optimization is crucial for the practical realization of quantum computing, where gate efficiency directly impacts the overall feasibility of quantum computations.

Once all multi-controlled gates have been decomposed into CNOTs and single-qubit gates, the next step in the compilation process is adapting the quantum circuit to the connectivity constraints of the quantum computer, a process known as circuit mapping. This process involves assigning an initial physical qubit to each qubit in the program and adding SWAP gates (which can be implemented using three CNOTs) to satisfy the connectivity constraints. This step of the compilation process often increases the CNOT count of the final circuit. Evaluating the impact of the proposed optimization--using an auxiliary qubit for decomposition--on circuit mapping remains an area for future work. However, we believe that the circuit mapping process is unlikely to negate the benefits of the proposed optimization, as reducing the number of CNOTs minimizes potential connectivity mismatches that would otherwise require additional SWAP gates.

Future work may focus on exploring the impact of this optimization in other quantum algorithms. Additionally, investigating the effect of this optimization in real-world quantum hardware scenarios will be essential to fully understand and harness its benefits. Note that the auxiliary qubit cannot be used in parallel; therefore, allocating more auxiliary qubits may be beneficial. Also, due to the connectivity of the qubits in the quantum computer, allocating more auxiliary qubits in strategic locations may positively improve the quantum circuit mapping. Another possible optimization is to decompose the multi-controlled Pauli gates using dirty auxiliary qubits, with the quantum compiler automatically selects the auxiliary qubits, which would reduce the decomposition of those gates to at most $4n$ CNOTs without impacting high-level quantum programming.

\section*{Acknowledgement}

EID acknowledges the Conselho Nacional de Desenvolvimento Científico e Tecnológico - CNPq through grant number 409673/2022-6 and ECRR acknowledges the Coordenação de Aperfeiçoamento de Pessoal de Nível Superior - CAPES, Finance Code 001.

\bibliographystyle{plainnat}
\bibliography{main}

\end{document}